\documentclass[12pt,cite,epsf,epsfig]{article}
\usepackage{epsfig}

\title{Finite Temperature Effects And Axion Cosmology}
\author{Namit Mahajan$^a$\thanks{E--mail :
    nmahajan@theory.tifr.res.in}~
and Sukanta Panda $^b$\thanks{E--mail: sukanta@delta.ft.uam.es}\\ \\
	{$^a$\em Department of Theoretical Physics,
	 Tata Institute of Fundamental Research,}\\ 
          {\em Homi Bhabha Road, Mumbai - 400005, India.}\\ \\
{$^b$\em Departamento de Fisica Teorica C-XI and Instituto de Fisica 
Teorica C-XVI}\\
{\em  Universidad Autonoma de Madrid, Cantoblanco, E-28049 Madrid, Spain. }}

\begin{document}

\maketitle

\begin{abstract}
We investigate the impact of finite temperature effects on axions in
the context of cosmology. The temperature dependence of the
decay constant is modeled analogous to pions. For the two
interesting cases considered here, we find that the temperature effects
do lead to changes relevant for detailed and precise abundance and
rate calculations. We also find that the axion decoupling temperature
starts showing large deviations for larger values of the axion decay
constant.\\ \\
\end{abstract}

\begin{flushright}
TIFR/TH/06-10\\
FTUAM-06-05\\IFT-UAM/CSIC-06-20
\end{flushright}

\newpage

\section{Introduction}
\par Axions, the pseudo Goldstone bosons, associated with the
Peccei-Quinn symmetry, $U(1)_{PQ}$, remain the most popular solution
to the strong CP problem \cite{pq1,pq2,pq3,pq4}. 
Axions, QCD type or some other variant, 
appear in many extensions of the standard model. The mass and coupling
of the axion is inversely proportional to the scale at which
$U(1)_{PQ}$ is broken, denoted by $f_a$, and here after referred to as
Peccei-Quinn (PQ) scale or axion decay constant interchangeably.
This has important implications for axion cosmology because axions can
only be produced thermally when the temperature of the universe drops
below $f_a$. The PQ scale, $f_a$, is constrained by astrophysics, 
cosmology and 
laboratory experiments. For a review of latest bounds on $f_a$ refer to 
\cite{sikivie,raffelt1}, while a more comprehensive review/collection
of (earlier) bounds on $f_a$ 
can be found in \cite{masso1,raffelt2,raffelt3,masso2,masso3}.
Further, axions are also one of the theoretically well motivated
contenders for the cold dark matter in the universe. The DAMA
collaboration has also explored the possibility of scalar and
pseudo-scalar interpretation of their results \cite{dama}. More recently,
results from the PVLAS experiment \cite{pvlas} indicate the existence
of an axion like particle, though at the moment it is difficult to
reconcile any standard axion like particle with
these results. For a summary of some of the possible ways out
to this difficulty refer to \cite{ringwaldpvlas}.

\par In this note we would like to study some cosmological
implications/effects in the context of axions. To the best of our
knowledge, all the existing studies have ignored the finite
temperature effects to axion decay constant 
while discussing issues like relic axion abundance etc. We intend to
explore the possible impact of such corrections on axion physics at
different epochs of the early universe. Before proceeding with the
finite temperature effects, we fix our notation. It is worthwhile to
mention that axions share the basic properties with the pions,
notably, the axion coupling to two gluons is given by
the anomaly term (the two photon coupling has an analogous form)
\begin{equation}
L =  \frac{1}{f_a} \frac{\alpha_s}{8 \pi} G^{b\mu\nu} 
{\tilde{G}}^{b}_{\mu \nu} a  \ ,
\end{equation}
where $\alpha_s$ is the QCD coupling constant and ${\tilde{G}}^{b}_{\mu 
\nu}$ is the dual of the gluon field defined as
\begin{equation}
{\tilde{G}}^{b}_{\mu \nu} = \frac{1}{2} \epsilon_{\mu \nu \rho \sigma} 
G^{b~\rho\sigma} \ ,
\end{equation}
where b is the color index. 
Let us rewrite the above as
\begin{equation}
L =  g_{agg}\epsilon_{\mu\nu\rho\sigma}G^{b\mu\nu} 
G^{b\rho\sigma} a  \ ,
\end{equation}
where $g_{agg}$ is the axion coupling which is inversely proportional to $f_a$.
 The axions can be further divided
into two categories - (a) Hadronic axions which couple to quarks only
and (b) non-Hadronic axions having a coupling to the leptons.
 Mass of hadronic axions is given by
\begin{equation}
m_a = \frac{z^{1/2}}{1+z} \frac{f_{\pi}m_{\pi}}{f_a} \ ,
\end{equation}
where $f_{\pi}=93$ MeV is the pion decay constant, and $z=m_u/m_d$ is 
the mass ratio of up and down quarks. $z$ can lie within the interval 
$0.3-0.7$ \cite{rpp}. Here we take $z=0.56$ \cite{gasser,leut}.

\par Having noted that the axion coupling to massless gauge bosons is
the same as pion-photon-photon coupling, we model the temperature
effects in an analogous way as well. To this end, we first note that
the results on chiral dynamics and anomaly at finite temperature
\cite{finitepion,tytgat} imply the following
\begin{equation}
f_{\pi}(T) = \left(1 - \frac{1}{12}\frac{T^2}{f_{\pi}^2}\right)f_{\pi} \ ,
\label{decay}
\end{equation}
 where $f_{\pi} \equiv f_{\pi}(T=0) \sim 93$ MeV
 is the zero temperature pion decay constant measured in the
 laboratory. It is important to keep in mind that such a result is
 valid only for $T < f_{\pi}$. In the case of
 axions, it is natural to look at temperatures lower than the axion
 decay constant as it is only then that the axions can be thermally
 produced. Therefore, the above result for the pions can be directly
 extended to the axion case. The analysis in \cite{tytgat} however
 brings out a crucial point. The effective pion-photon-photon coupling
 is inversely proportional to the decay constant and one would have
 expected that the coupling changes the way dictated by the change in
 decay constant with the temperature, i.e., one would have expected the
 coupling to change as $g_{\pi\gamma\gamma} \to
 g_{\pi\gamma\gamma}\left(1 -
 \frac{1}{12}\frac{T^2}{f_{\pi}^2}\right)^{-1}$ or that to
 ${\mathcal{O}}(T^2/f_{\pi}^2)$, $ g_{\pi\gamma\gamma}(T) \sim
 1/f_{\pi}(T)$. What is found instead is that to this order $
 g_{\pi\gamma\gamma}(T) \sim f_{\pi}(T),$ i.e.,
\begin{equation}
g_{\pi\gamma\gamma}(T) = \left(1 -
\frac{1}{12}\frac{T^2}{f_{\pi}^2}\right)g_{\pi\gamma\gamma}. 
\label{coupling} 
\end{equation}
We therefore take this form for the axion-gluon-gluon coupling and
work with this.\footnote{It is worthwhile to point out that the
  configuration considered in \cite{tytgat} does not corespond to the
  physical situation while the authors in 
\cite{sourendu} consider the decay of pseudoscalar to real photons and
find difference with the results obtained in \cite{tytgat}. See
\cite{gelis} for a discussion about this difference and the
resolution. However, we only use the results concerning the change in
decay constant with temperature and therefore this issue does not
affect us here.}

 Further, following \cite{finitepion,pisarski}, we have for the
pion mass (again in the regime $T < f_{\pi}$),
\begin{equation}
m_{\pi}(T) = m_{\pi} \left(1 + \frac{T^2}{24 f_{\pi}^2}\right).
\label{mass}
\end{equation}
Note that since the
limits on axion mass indicate a very small mass, we do not consider
temperature dependent effects for axion mass. 

\par We are now ready to explore two epochs in the history
of the universe, particularly important and relevant for axion
cosmology. First, we study axion thermalization and abundance in
the very early universe including the temperature effects. The second
case investigated is the axion hadron interactions in the post QCD
era. In both these cases we find that the temperature effects - via
axion or pion decay constant, pion mass or axion-gluon-gluon coupling,
whichever is applicable at the relevant epoch - do lead to change in
the results for certain values of the temperatures.

\section{Axion thermalization and abundance}
\begin{figure}[ht]
\begin{center}
\epsfig{file=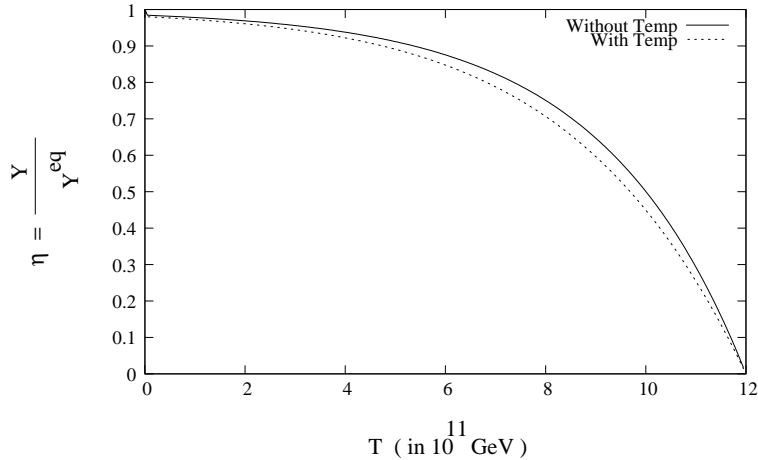,width=10cm}
\end{center}
\caption{Abundance of axions for $f_a=1.2 \times 10^{12} GeV.$}
\label{fig1}
\end{figure}

As mentioned earlier, axions can be thermally produced in the universe
once the temperature falls below the axion decay constant. However,
axions can also be produced via some non-thermal processes (see
\cite{mas} and references therein). In
\cite{masso4}, the authors identify the conditions under which there
is significant production and/or thermalization of axions. 
We therefore closely follow \cite{masso4} but now include temperature
effects as mentioned before. The main thermalization processes are 
axion-gluon and axion-quark scatterings. From Eq.(\ref{decay}) and 
Eq.(\ref{coupling}), we see that both $f_a$ and $g_{agg}$ decrease
with the inclusion of the temperature effects. This means that the
axions can be thermally produced at a slightly lower temperature and
the interaction rate also goes down a bit.
The Boltzmann equation for the abundance 
in an expanding universe is given by 
\begin{equation}
x \frac{dY}{dx} = \frac{\Gamma}{H}(Y^{eq} - Y) \ ,
\label{abundance}
\end{equation}
where $x=\frac{f_a}{T},$ $H$ is the Hubble expansion rate and $\Gamma$ 
is the thermally averaged rate of 
reaction of axion processes. 
In the radiation dominated era, the Friedmann equation yields
\begin{equation}
H = \left(\frac{4 \pi^3 g_{eff}}{45}\right)^{1/2}\frac{T^2}{M_P} \ ,
\end{equation}
where $g_{eff}$ is the effective degrees of freedom at the temperature
$T$. For a generic process $a + i \rightarrow j + k$ ($i,j,k \neq
a$), thermally averaged rate is
\begin{equation}
\Gamma =
\frac{1}{n_a}\int~\tilde{dp_a}\tilde{dp_i}\tilde{dp_j}\tilde{dp_k} 
(2\pi)^4\delta^4(p_a+p_i-p_k-p_k)\vert{\mathcal{M}}\vert^2~f_af_i(1+f_j)(1+f_k) 
\ ,
\end{equation}
where $\tilde{dp}=\frac{d^3p}{(2\pi)^32E}$, $f$'s are the
  equilibrium distribution functions and $n_a$ is the equilibrium 
axion number density. Compared to the expression for the
thermal averaged interaction rate in \cite{masso4}, we will have an
extra temperature dependent factor stemming from the correction to the
axion-gluon-gluon coupling and the rate now looks
\begin{equation}
\Gamma \simeq  7.1 \times 10^{-6} 
\frac{T^3}{f_a^2}\left(1- \frac{T^2}{12 
f_a^2}\right)^2 =\Gamma_0 {\left(1- \frac{T^2}{12 
f_a^2}\right)^2} \ . 
\end{equation}
$Y^{eq}$ is the
abundance when axions are in thermal equilibrium with other SM
particles. In terms of variables $\eta = \frac{Y}{Y^{eq}}$ and $k= x 
\frac{\Gamma_0}{H},$ Eq.(\ref{abundance}) becomes
\begin{equation}
x^2\frac{d\eta}{dx} = k\left(1-\frac{1}{12 x^2}\right)^2(1 - \eta)\ .
\label{abundance1}
\end{equation}
Retaining only the leading term, the solution is
\begin{equation}
\eta(x) = 1 + C~exp\left[\frac{k}{x} - \frac{k}{18 x^3}\right] \ .
\end{equation}
where $C$ is a constant determined by boundary condition, 
$\eta=0$ for $\frac{f_a(T)}{T}=1$. This gives $\eta=0$ at $x\sim 1.08$ 
instead of at $x=1$ as the boundary condition and the constant $C$ is
evaluated accordingly. Fig.\ref{fig1} compares the axion abundance with and
without the temperature effects taken into consideration for $f_a =
1.2 \times 10^{12}$ GeV. We see that
there is a perceptible change in the abundance for the temperature
range ($6-10$)$\times 10^{11}$ GeV. As a consequence of the shift in
the $x$ value from unity to $1.08$, the results in \cite{masso4}
shift by this factor.

\section{Hadronic axion in the post QCD era}
\par The hadronic axion has no lowest level coupling to the charged leptons
and therefore the induced coupling is expected to be small enough to
be neglected altogether. For updated bounds on hadronic axion in this context
refer to \cite{hann}. Although the exact limits on the mass and decay
constant of the axion are model dependent, up to order unity factors,
the limits from various sources suggest: $f_a \geq 0.6 \times 10^9$
GeV and $m_a \leq 0.01$ eV. Stellar energy loss constraints suggest that
in the post QCD era, the axion coupling to electrons and photons are
not significant.  The axion coupling to nucleons is constrained by
the observed  
neutrino signal from Supernovae 1987A \cite{raffelt2,eidel}. Supernovae
observations give stringent constraints on the mass and coupling:
$m_a \geq 0.01$ eV or $f_a \leq0.6 \times 10^9$ GeV. However it still 
leaves a domain for hadronic axion where $f_a$ lies in the range $3 \times 
10^5$ GeV and $3 \times 10^6$ GeV \cite{chang, moroi}. But this range is 
disfavored by a detailed combined analysis of hadronic axion with 
cosmological data like large scale structure, cosmic microwave 
background, supernova luminosity distances, Lyman-$\alpha$ forest and 
Hubble parameter \cite{hann}, which obtains the following limits:
$m_a < 1.05$ eV or equivalently $f_a > 5.7 \times 10^6$ GeV. However,
in the post QCD era of interest to us here, the number density of the
nucleons falls much faster than that of pions and therefore, nucleons
can be effectively neglected compared to the pions. That the nucleon
contribution to the thermalization and decoupling of axions in this
era is much smaller than the pion contribution has been explicitly
verified in \cite{chang}.

\par The axion-pion interaction is of the form \cite{chang}
\begin{equation}
L_{a \pi} = \frac{C_{a \pi}}{f_{\pi} f_a} \left( \pi^0 \pi^+ 
\partial_{\mu}\pi^- + \pi^0 \pi^- \partial_{\mu}\pi^+ 
- 2 \pi^+ \pi^- \partial_{\mu}\pi^0 \right) \partial_{\mu}a \ .
\end{equation}
In hadronic axion models, the coupling constant is
\begin{equation}
C_{a \pi} = \frac{1-z}{3 (1+z)} \ .
\end{equation}

\par
\begin{figure}[ht]
\begin{center}
\epsfig{file=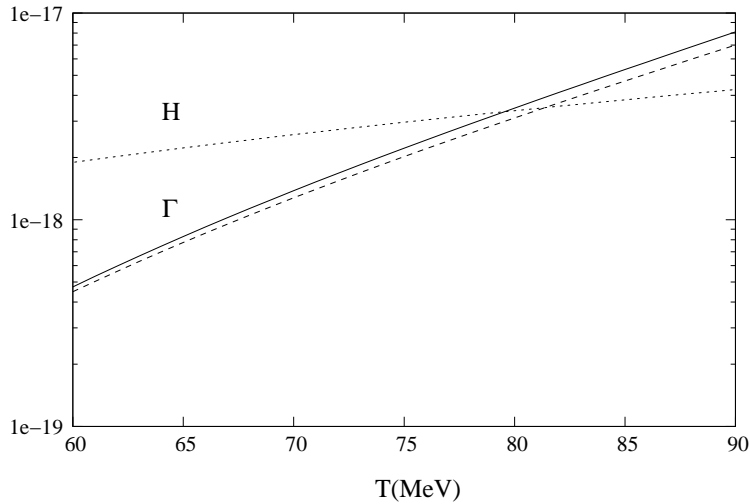,width=10cm}
\end{center}
\caption{Hadronic axion reaction rate with(solid) and without(dashed) 
temperature effect for $f_a=10^7 GeV.$}
\label{fig2}
\end{figure}

  The relevant processes are $a \pi^{\pm}\rightarrow \pi^0
\pi^{\pm}$ and $a \pi^0\rightarrow \pi^+\pi^-$. Following
\cite{hann,chang}, we calculate the axion decoupling temperature in
this context but now taking the finite temperature effects in the pion
decay constant and mass into account given in Eq.(\ref{decay}) and
Eq.(\ref{mass}) respectively (valid for $T<f_{\pi}$). The decoupling
temperature is that where the Hubble expansion rate equals the
thermally averaged interaction rate. In the Fig.\ref{fig2}, we plot
the expansion rate of the universe ($H$) and the interaction rates
($\Gamma$) - with and without temperature effects put in. The points
where the $\Gamma$ curves cut the $H$ curve give the decoupling
temperature. This is shown in the figure for $f_a=10^7$ GeV.
It is clear from the figure that the decoupling temperature is lowered
 once the finite temperature effects are included and the reason is
 easy to see. In the epoch of interest, the temperatures are much lower
 than the axion decay constant and therefore, we do not consider any
 temperature dependent effects in $f_a$, while $f_{\pi}$ decreases and
 $m_{\pi}$ increases with temperature. Both these effects lead to
 larger interaction rate and therefore, the axions decouple later in
 time or at lower temperatures. From the figure, it is also clear that
 the difference in the two cases is not very large.
 We also compare decoupling temperature of axions 
in the present calculation with the previous ones (without the
temperature effects) for several representative 
values of $f_a$ in Table \ref{t}. We find that for higher values of
 $f_a$, the temperature effects start becoming significant. However,
 within the approximation adopted here, it is not possible to explore
 the range $f_a \geq 1.5\times 10^7$ GeV as the temperatures exceed the
 pion decay constant, thus taking us away from the validity of the
 approximation employed. It is therefore important to investigate the
 temperature effects going beyond this approximation and exploring the
 consequences. 
\begin{table}[h]
\begin{center}
\begin{tabular}{|l|l|l|l|r|}
\hline
$f_a$(GeV) & $T_{D1}$(MeV) & $T_{D2}$(MeV)\\
\hline
3 $\times 10^5$ & 26.43 & 26.43\\
1 $\times 10^6$ & 35.34 & 35.34\\
3 $\times 10^6$ & 49.84 & 49.5\\
1 $\times 10^7$ & 81.04 & 79.12\\
1.2 $\times 10^7$ & 87.61 & 85.48\\
1.3 $\times 10^7$ &  90.1 & 87.9\\
\hline
\end{tabular}
\caption{Decoupling temperature of axions $T_{D1}$ (without)
 and $T_{D2}$ (with) temperature effects for different 
values of $f_a.$}
\label{t}
\end{center}
\end{table}

\section{Discussion}
\par We have investigated the impact of temperature effects, though in
a limited sense, on the axion cosmology. We have focused our
attention of the temperature effects in the axion decay constant,
pion decay constant and pion mass in two important epochs of the
universe. The results show that there can be perceptible differences
in the predictions of abundances and decoupling temperatures. However,
we must remark that these effects have been considered in the
approximation when the temperature of the universe is lower than the
decay constant(s). In the case of axion thermalization in the early
universe this has a physical meaning because it is only then that the
axions can be thermally produced. In this case we see some change in
the abundance of the axions with temperature. In the case of axion
interacting with pions in the post QCD era, we find that temperature
effects in the pion mass and decay constant lead to a lower decoupling
temperature for the axions. We find that as the axion decay constant
is increased the difference between the zero temperature and finite
temperature cases starts becoming significant.
However, we have worked in the limit when
temperatures are smaller than $f_{\pi}$. This puts a natural limit to
the region that we can explore. It is worthwhile to try to
explore what happens when $T>f_{\pi}$. This will require going beyond
the validity of the present calculation. We would also like to mention
that a more detailed and consistent calculation requires computing the
matrix elements including thermal loops
also to the same order in temperature corrections and
then studying the full impact.

\par In summary, we can say that we have studied the impact of
temperature effects due to a very limited source(s) 
entering the axion cosmology. The results show perceptible departure from the
zero temperature calculations. Let us just compare the situation with
the case of neutrino decoupling. In that case, naively, the temperature effects
shift the decoupling temperature by a seemingly small amount. However,
for precision studies, such a shift makes a significant difference and
hence turns out to be important. Our approach is also guided by this
fact. See \cite{subir} (and references therein) for related discussion.
Whether, finite temperature effects are
actually significant and substantial can be decided by more detailed
computations, taking into account all sources of corrections to this
order. Also, in the case of axion decoupling in the post QCD era,
it'll be interesting to try to go beyond the approximation of
$T<f_{\pi}$ and include finite temperature effects in the masses and
all other parameters entering the expressions. This can have serious
implications for axion cosmology and can lead to very different
bounds/limits on various parameters, thereby possibly changing our
current understanding.

\vskip 1cm
{\bf Acknowledgments}\\
The work of S.P is supported by the Ministerio de Educacion y Ciencia, 
Spain.


\begin{thebibliography}{unsrt}

\bibitem{pq1}
  R.~D.~Peccei and H.~R.~Quinn,
  ``CP Conservation in the presence of pseudoparticles'',
  Phys.\ Rev.\ Lett.\  {\bf 38} (1977) 1440.

\bibitem{pq2}
  R.~D.~Peccei and H.~R.~Quinn,
  ``Constraints imposed by CP conservation in the presence of
  pseudoparticles'',
  Phys.\ Rev.\ D {\bf 16} (1977) 1791.

\bibitem{pq3}
  S.~Weinberg,
  ``A new light boson?','
  Phys.\ Rev.\ Lett.\  {\bf 40} (1978) 223.

\bibitem{pq4}
  F.~Wilczek,
  ``Problem of strong P and T invariance in the presence of
  instantons'',
  Phys.\ Rev.\ Lett.\  {\bf 40} (1978) 279.



\bibitem{sikivie}
  P. Sikivie,
  ``Axions 05'',
  [hep-ph/0509198].

\bibitem{raffelt1}
  G. G. Raffelt,
  ``Axions: recent searches and new limits'',
  [hep-ph/0504152].

\bibitem{masso1}
  Eduardo Masso,
  ``Axions'',
 [hep-ph/0312064]; Nucl.\ Phys.\ Proc.\ Suppl. {\bf 114} (2003) 67
 [hep-ph/0209132].

\bibitem{raffelt2}
  G.~G.~Raffelt,
  ``Particle physics from stars'',
  Annu.\ Rev.\ Nucl.\ Part.\ Sci.\  {\bf 49} (1999) 163
  [hep-ph/9903472].

\bibitem{raffelt3}
  G.~G.~Raffelt,
  ``Stars as laboratories for fundamental physics: The astrophysics 
of neutrinos, axions and other weakly interacting particles'', Chicago, 
USA: University press (1996) 664 p.

\bibitem{masso2}
  Eduardo Masso and Ramon Toldra,
  ``On a light spinless particle coupled to photons'',
Phys.\ Rev.\ D {\bf 52} (1995) 1755
 [hep-ph/9503293].

\bibitem{masso3}
  Eduardo Masso and Ramon Toldra,
  ``New constraints on a light spinless particle coupled to photons'',
Phys.\ Rev.\ D {\bf 55} (1997) 7967
 [hep-ph/9503293].

\bibitem{dama}R.~Bernabei et al.,
`` Investigating pseudoscalar and scalar dark matter'',
  Int. J. Mod. Phys. A{\bf 21} (2006) 1445. 


\bibitem{pvlas}E.~Zavattini et. al, PVLAS Collab.,
``PVLAS: probing vacuum with polarized light'', [hep-ex/0512022]; 
`` Experimental observation of optical rotation generated in vacuum by
  a magnetic field'', [hep-ex/0507107].


\bibitem{ringwaldpvlas}
 A.~Ringwald,
``Axion interpretation of the PVLAS data?'', [hep-ph/0511184].




\bibitem{rpp}
  S.~Eidelman {\it et al.}  (Particle Data Group),
  ``Review of particle physics'',
  Phys.\ Lett.\ B {\bf 592} (2004) 1.


\bibitem{gasser}
  J.~Gasser and H.~Leutwyler,
  ``Quark masses'',
  Phys.\ Rept.\ {\bf 87} (1982) 77.

\bibitem{leut}H.~Leutwyler,
  ``The ratios of the light quark masses'',
  Phys.\ Lett.\ B {\bf 378} (1996) 313
  [hep-ph/9602366].

\bibitem{finitepion}J.~Gasser and H.~Leutwyler,
``Light quarks at low temperatures'',
Phys. \ Lett. \ B {\bf 184} (1987) 83;
A.~Bochkarev and J.~I.~Kapusta,
`` Chiral symmetry at finite temperature: Linear versus nonlinear
sigma models'', Phys. \ Rev. \ D {\bf 54} (1996) 4066;
D.~Toublan,
``Pion dynamics at finite temperature'',
Phys. \ Rev. \ D {\bf 56} (1997) 5629.


\bibitem{tytgat}
R.~D.~Pisarski, T.~L.~Trueman and M.~H.~G.~Tytgat,
``How $\pi^0\to\gamma\gamma$ changes with temperature'',
Phys. \ Rev. \ D {\bf 56} (1997) 7077.


\bibitem{sourendu}S.~Gupta and S.~N.~Nayak,
``Thermal effects on two-photon decays of pseudo-scalars'',
  [hep-ph/9702205].

\bibitem{gelis}F.~Gelis,
``Ambiguities in the zero momentum limit of the thermal
  $\pi^0\gamma\gamma$ triangle diagram'',
Phys. \ Rev. \ D {\bf 59 (1999)} 076004-1.


\bibitem{pisarski} Robert D. Pisarski and Michel Tytgat, "Propagation 
of cool pions", Phys. Rev. D {\bf 54} (1996) R2989.

\bibitem{mas} Eduard Masso, Francesc Rota, Gabriel Zsembinszki, 
"Planck-scale effects on global symmetries: Cosmology of 
Pseudo-Goldstone Bosons", Phys.Rev. D {\bf 70} (2004)
115009 [hep-ph/0404289].


\bibitem{masso4} Eduard Masso, Francesc Rota, Gabriel Zsembinszki, "On 
axion thermalization in the early universe", Phys.Rev. D {\bf 66} (2002) 
023004 [hep-ph/0203221].

\bibitem{hann} S.Hannestad, A.Mirizzi and G.Raffelt, 
  "New cosmological 
mass limit on thermal axions", JCAP {\bf 0507} (2005) 002 
[hep-ph/0504102].



\bibitem{chang}
  S.~Chang and K.~Choi,
  ``Hadronic axion window and the big bang nucleosynthesis'',
  Phys.\ Lett.\ B {\bf 316} (1993) 51

\bibitem{moroi}
  T.~Moroi and H.~Murayama,
  ``Axionic hot dark matter in the hadronic axion window'',
  Phys.\ Lett.\ B {\bf 440} (1998) 69
  [hep-ph/9804291].

\bibitem{zhit}
  A.~R.~Zhitnitsky,
  ``On possible suppression of the axion hadron interactions'',
  Sov.\ J.\ Nucl.\ Phys.\  {\bf 31} (1980) 260
  [Yad.\ Fiz.\  {\bf 31} (1980) 497].



\bibitem{eidel} S. Eidelman et al [Particle Data Group], "Review of 
particle physics", Phys. Lett. B {\bf 592} (2004) 1.

\bibitem{subir}S.~Sarkar,
`` Big bang nucleosynthesis and physics beyond the standard model'', 
Rept. Prog. Phys. {\bf 59} (1996) 1493. 

\end{thebibliography}
\end{document}